\documentstyle[12pt]{article} 
\def\F{{\bf F}}
\def\A{{\bf A}}
\def\J{{\bf J}}

\def\beq{\begin{equation}}
\def\eeq{\end{equation}}
\def\pl{\partial}

\def\al{\alpha}
\def\bt{\beta}

\def\ga{\gamma}
\def\de{\delta}
\def\De{\Delta}

\def\Si{\Sigma}
\def\te{\theta}
\def\La{\Lambda}
\def\lam{\lambda}

\def\om{\omega}
\def\ep{\epsilon}

\def\sq{\sqrt}

\def\l{\left (}
\def\r{\right )}
\def\fr{\frac}
\def\la{\label}
\def\hs{\hspace}
\def\vs{\vspace}
\def\inf{\infty}
\def\ran{\rangle}
\def\lan{\langle}
\def\ov{\overline}

\def\tm{\times}

\begin{document}

\begin{titlepage}
\begin{flushright}
BA-02-24\\
\end{flushright}

\begin{center}
{\Large\bf    
Gauge Origin of Baryon Number Conservation and\\
\vspace{0.2cm}
Suppressed Neutrino Masses 
From Five Dimensions}
\end{center}
\vspace{0.5cm}
\begin{center}
{\large 
Chin-Aik Lee$^{a}$
\footnote {E-mail address: jlca@udel.edu},~
{}~Qaisar Shafi$^{a}$
\footnote {E-mail address: shafi@bartol.udel.edu},~ 
{}~Zurab Tavartkiladze$^{b, c}$
\footnote {E-mail address: Z.Tavartkiladze@ThPhys.Uni-Heidelberg.DE} 
}
\vspace{0.5cm}

$^a${\em Bartol Research Institute, University of Delaware,
Newark, DE 19716, USA \\

$^b$ Institute for Theoretical Physics, Heidelberg University, 
Philosophenweg 16, \\
D-69120 Heidelberg, Germany\\

$^c$ Institute of Physics,
Georgian Academy of Sciences, Tbilisi 380077, Georgia\\
}

\end{center}
\vspace{1.0cm}

\begin{abstract}

We consider a 5D SUSY $SU(3)_c\times SU(2)_L\times 
U(1)_Y\times {\cal U}(1)$ model compactified 
on an $S^{(1)}/Z_2$ orbifold. To cancel anomalies arising from the
presence of 
${\cal U}(1)$, we employ a Chern-Simons 
term and also 
chiral fields which could reside on the brane or in the bulk depending on
the 
model. The presence of ${\cal U}(1)$ symmetry leads to baryon number 
conservation, gives rise to matter parity, 
and permits satisfactory neutrino masses and mixings 
even for a low fundamental scale. The brane Fayet-Iliopoulos D-terms 
naturally break 
${\cal U}(1)$, leaving
$N=1$ SUSY unbroken in 4 dimensions.

\end{abstract}

\end{titlepage}

\section{Introduction}

The so far unsuccessful search for proton decay by the SuperKamiokande
experiment \cite{SKpdecay} 
has yielded a lower bound of around $10^{33}$ years on the lifetime, which
proves 
especially challenging for supersymmetric models that allow the decay to
proceed
via the dimension five operators
\beq
\fr{\lam }{M}qqql~~~~~{\rm and}~~~~~\fr{\lam'}{M}u^cu^cd^ce^c~,
\la{intd5}
\eeq
where $M\sim M_{\rm Pl}=2.4\cdot 10^{18}$~GeV denotes reduced
Planck mass. The dimensionless parameters $\lam,
\lam'$ must be $<10^{-8}$ or so, which demands some reasonable
explanation.
The suppression of such $d=5$
operators can be realized by either imposing discrete gauge
\cite{gaugeZ}, flavor
\cite{d5flavor}, string-induced anomalous $U(1)$ \cite{stringu1} or
$R$-symmetries \cite{d5par}.
One must also suppress $d=5$
operators emerging through the exchange of additional
states,
such as the colored triplets appearing in GUTs. Various
mechanisms can be applied \cite{stringu1}-\cite{d5sup2} to
this
end, making the nucleon sufficiently long 
lived\footnote{Let us note that within SUSY GUTs, the decays mediated 
by the $X$ and $Y$
gauge bosons are adequately suppressed if the GUT
scale, $M_G\sim 2\cdot 10^{16}$~GeV.}.

The problem of B conservation becomes much more acute
in extradimensional theories
with a low fundamental scale. The main phenomenological motivation for
these kinds of models is the possibility of resolving the gauge hierarchy
problem \cite{tevgr}. However, lowering the fundamental mass scale $M_f$
down
to a few TeV increases the $d=5$ operator induced nucleon decay
amplitude
by a factor of $M_{\rm Pl}/M_f\sim 10^{16}$ unless some additional
mechanism
for B conservation is applied. 
In ref. \cite{gaugeB}, scenarios with gauged baryon
number were considered and
the matter sector was extended in order to cancel the anomalies.
Ref. \cite{branesupr} suggested scenarios in
which quarks and leptons are localized on different 3-branes
separated in the extra dimension(s). As a result, baryon number
violating operators can be strongly suppressed.
In ref. \cite{5DSU5}, within the framework of a five dimensional
(5D) $SU(5)$
orbifold GUT, certain $d=5$
operators were eliminated using
special prescriptions of orbifold symmetry
parities. It is also possible in such models to obtain GUT
symmetry
breaking and doublet-triplet splitting.
However, Planck scale $d=5$ operators can still be problematic and 
additional care must be taken to suppress them \cite{4-4}.
 
In this paper, we present a new scenario in which 
baryon number arises as an accidental symmetry at
the 4D level, which originates from  5D  
SUSY
$SU(3)_c\times SU(2)_L\times U(1)_Y$, supplemented with a
${\cal U}(1)$ symmetry.
After imposing a $Z_2$ projection, ${\cal U}(1)$ becomes anomalous on the 
fixed points. The 4D ${\cal U}(1)^3$ anomaly is cancelled by
a bulk Chern-Simons (CS) term \cite{CSterm},
\cite{susyCS}-\cite{nilles}. 
The known quark, lepton and higgs
superfields carry non-trivial ${\cal U}(1)$ charges,
whereas their $N=2$ mirrors carry opposite
charges. The mixed anomalies are cancelled through suitable assignments of
${\cal U}(1)$ charges for the quark-lepton superfields and by some
additional
chiral states.
 In the 5D bulk, we have a manifestly vectorlike
theory. After imposing a $S^{(1)}/Z_2$ orbifold compactification, we
obtain
4D $N=1$ SUSY $SU(3)_c\times SU(2)_L\times U(1)_Y$
supplemented with a ${\cal U}(1)$ gauge
factor. The latter is crucial not only for suppressing B
violating operators to the desired level, but also for
 obtaining appropriately suppressed neutrino masses and
automatic matter parity.
All this can
be achieved for various values of the fundamental mass scale,
with $M_f$ as low as $\sim 100$~TeV.
The ${\cal U}(1)$ symmetry can also be successfully employed as 
a flavor symmetry to explain  the hierarchies 
among the charged fermion masses
and their mixings.

\section{5D SUSY $SU(3)_c\times SU(2)_L\times U(1)_Y\times {\cal U}(1)$\\
on an $S^{(1)}/Z_2$ Orbifold}

It is  well known
fact that after a $Z_2$ projection,
bulk fermion fields can introduce an anomaly localized on both fixed
points \cite{indanom}, \cite{CSterm}, \cite{susyCS}-\cite{nilles}. This
anomaly can be written in the form
$D_AJ^{a,A}(y)=Q^a(y)f(y)$, where $Q^a\equiv
\frac{g^2}{32\pi^2}\ep^{\al \bt \ga \de }F^b_{\al \bt }
F^c_{\ga \de }{\rm Tr}[T^a\{T^b, T^c\}]$. 
Provided $\int^{2\pi R}_0 dy f(y)=0$, a bulk CS term can be added to
cancel the anomalies from the fermions. As shown in
refs. \cite{susyCS}-\cite{nilles},
$f(y)=\frac{1}{2}\left[\delta(y)+\delta(y-\pi R)\right]$, and
the integral of $f$ is
nonzero, which means we cannot cancel the 
anomaly simply with a CS term. However, this can be remedied by
adding
additional fermion fields in such a way that the integral of $f$ is zero.
 After this cancellation, the quantized theory will be free of 
local gauge anomalies\footnote{In string theories, anomaly cancellation
can occur through the
Green-Schwarz mechanism \cite{GSmech}. This also can be 
used efficiently
for baryon number conservation \cite{stringu1}.}.
Anomaly cancellation by adding a bulk
CS term was considered in refs. 
\cite{indanom}, \cite{CSterm}, \cite{susyCS}-\cite{nilles}. 
Here, we will exploit it
for obtaining baryon number conservation in four 
dimensions\footnote{This mechanism of 
anomaly cancellation was applied in ref. \cite{PQ} to 
gauge Peccei-Quinn $U(1)_{\rm PQ}$ symmetry in the bulk.}.
In 5D, we will introduce a ${\cal U}(1)$ gauge symmetry which, prior to
the addition of a CS term, is anomalous and suppresses dangerous
baryon number violating operators to the desired level.

Consider then a 5D supersymmetric $SU(3)_c\times SU(2)_L\times U(1)_Y$ 
supplemented with a ${\cal U}(1)$ gauge symmetry. In 4D notation,
the $N=2$ gauge superfield ${\bf V}_{N=2}=(V, \Phi )$ contains an
$N=1$ gauge superfield $V$ and a chiral superfield $\Phi $, both of which are 	
in the 
adjoint
representation of the gauge group.
The chiral
supermultiplet
${\bf H}_{N=2}=(H, \ov{H})$ contains two $N=1$ chiral
superfields
$H$ and $\ov{H}$ transforming as $p$ and $\ov{p}$-plets respectively
under the gauge group.  $H$ denotes all the `matter' and/or `scalar'
superfields of MSSM, while $\ov{H}$ denotes their mirrors.
In $N=1$ notation, the 5D action includes
\cite{susyCS}, \cite{5Dsusyact}:
\beq
S^{(5)}=\int d^5x({\cal L}^{(5)}_V+{\cal L}^{(5)}_{H})
\la{5dact},
\eeq
where
\beq
{\cal L}^{(5)}_V=\fr{1}{4g^2}\int d^2\te W^{\al }W_{\al }+{\rm h.c.}+
\fr{1}{g^2}\int d^4\te \l (\sq{2}\pl_5V+\Phi^{+})e^{-V}
(-\sq{2}\pl_5V+\Phi )e^V+
\pl_5e^{-V}\pl_5e^V\r~,
\la{lagv}
\eeq
\beq
{\cal L}^{(5)}_{H}=\int d^4\te \l H^{+}e^{-V}H +
\ov{H}e^{V}\ov{H}^{+} \r+
\int d^2\te \ov{H}\l M_{H}+\pl_5-\fr{1}{\sq{2}}\Phi \r H+
{\rm h.c.},
\la{lagf}
\eeq
and $W_{\alpha}$ are the supersymmetric field strengths.
The action in eq. (\ref{5dact}) is invariant under the gauge
transformations
$$
e^V\to e^{\La}e^Ve^{\La^{+}}~,~~~~
\Phi \to e^{\La }(\Phi -\sq{2}\pl_5)e^{-\La }~,
$$
\beq
H \to e^{\La }H~,~~~~~~~~\ov{H} \to \ov{H}e^{-\La }~.
\la{n2gaugetr}
\eeq 
In eqs. (\ref{lagv})-(\ref{n2gaugetr})
\beq
\Phi =\fr{1}{\sq{2}}(\Si+iA_5)+\sq{2}\te \psi +\te \te F~,
\la{defPhi}
\eeq
where $A_5$ is fifth component of a 5D gauge field and $\Si $ is the real
adjoint coming from 5D $N=1$ gauge supermultiplet.

We consider compactification on an $S^{(1)}/Z_2$ orbifold, with all
fields 
having a definite $Z_2$ parity.
States with positive and negative parities $H_{+}$,
$\ov{H}_{-}$ can be expressed as
$$
H_{+}=\fr{\sq{2}}{\sq{\pi R}}\sum_{n=0}^{n=\inf }H^{(n)}(x)\eta^{(n)}
\cos \l \fr{ny}{R}\r~,
$$
\beq
\ov{H}_{-}=\fr{\sq{2}}{\sq{\pi R}}\sum_{n=1}^{n=\inf }\ov{H}^{(n)}(x)
\sin \l \fr{ny}{R}\r~,
\la{expan1}
\eeq
where $\eta^{0}=1/\sq{2}$ and $\eta^{(n)}=1$ for $n\neq 0$.
As can be seen from eq. (\ref{expan1}), $\ov{H}_{-}$ does not have a zero 
mode. The fixed point $y=0$ is identified as
the 3-brane corresponding to our 4D world.

In 5D, 
we also introduce a SM singlet
superfield ${\bf X}_{N=2}=(X ,~\ov{X})$ which carries a
${\cal U}(1)$ charge and is crucial for ${\cal U}(1)$ symmetry breaking in
4D.
The field content of the 5D model is given by 
$$
Q_{N=2}=(q ,~\ov{q})~,~~~~U^c_{N=2}=(u^c, ~\ov{u}^c)~,~~~~
D^c_{N=2}=(d^c ,~\ov{d}^c)~,
$$
\beq
L_{N=2}=(l ,~\ov{l})~,~~~~~~~E^c_{N=2}=(e^c ,~\ov{e}^c)~,
\la{N2matter}
\eeq
\beq
H^u_{N=2}=(h_u, ~\ov{h}_u)~,~~~H^d_{N=2}=(h_d, ~\ov{h}_d)~,~~~
{\bf X}_{N=2}=(X ,~\ov{X})~.
\la{N2scalar}
\eeq
The ${\cal U}(1)$ charges and $Z_2$ parities  of the various
components
of the gauge 
(${\bf V}_{N=2}$)
and `matter'/`scalar' (${\bf H}_{N=2}$) superfields are displayed
in Table \ref{t:scalar}. Note that $a$, $b$, $\al$ and $\ga$ are numbers to
be specified
later and $n$ is a positive integer. 
%

\begin{table}
\caption{The ${\cal U}(1)$ charges and $Z_2$ parities of gauge, matter
and scalar superfields}

\label{t:scalar}
$$\begin{array}{|c|c|c|}

\hline
N=2~ {\rm Supermultiplet} &{\cal U}(1)~ {\rm Charge} &
Z_2~{\rm Parity} \\
\hline

{\rm All}~{\bf V}_{N=2}=(V, ~ \Phi )  & (0, ~0) &(+, ~-) \\
{\bf X}_{N=2}=(X, ~\ov{X}) & (1, ~ -1)  &(+, ~-) \\
 Q_{N=2}=(q, ~\ov{q}) & (a, ~-a)  &(+, ~-) \\
 U^c_{N=2}=(u^c ,~ \ov{u}^c )  & (-a+\al , ~a-\al ) &(+, ~-) \\
 D^c_{N=2}=(d^c ,~ \ov{d}^c ) & (-a-n+\ga , ~a+n-\ga )  &(+, ~-) \\
 L_{N=2}=(l ,~\ov{l}  ) & (b+\ga , ~-b-\ga )  &(+, ~-) \\
 E^c_{N=2}=(e^c ,~ \ov{e}^c ) & (-b-n , ~b+n )  &(+, ~-) \\
 H^u_{N=2}=(h_u ,~ \ov{h}_u  ) & (-\al  , ~\al )  &(+, ~-) \\
 H^d_{N=2}=(h_d ,~  \ov{h}_d ) & (-\ga , ~\ga )  &(+, ~-) \\

\hline

\end{array}$$

\end{table}


After projecting out states with negative $Z_2$ parity, we
effectively have 4D N=1 MSSM supplemented with a ${\cal
U}(1)$ gauge
symmetry and a superfield $X$. In the next section we shall see that the 
${\cal U}(1)^3$
anomaly from the fermions can be cancelled
by a compensating
contribution from a CS action involving the ${\cal U}(1)$ gauge
field.

\section{Anomaly Cancellation}

The 5D anomaly from bulk fermion fields on
$S^{(1)}/Z_2$ is given by
\cite{indanom}, \cite{CSterm}, \cite{susyCS}-\cite{nilles}
\begin{equation}
D_AJ^{a,A}(y)=\frac{Q^a(y)}{2}\left[\delta(y)+\delta(y-\pi R)\right]
\end{equation}
where D is the covariant derivative, $A$ denotes the five spatial
dimensions, $a$ labels the $T^a$
generator of the gauge group,
and
\begin{equation}
Q^a\equiv\frac{g^2}{32\pi^2}\epsilon^{\al \bt \ga \de }
F^b_{\al \bt }F^c_{\gamma\delta}{\rm Tr}[T^a\{T^b, T^c\}]~.
\end{equation}
By contrast, the
anomalies due to brane fermion fields localized on the $y=0$ or the
$y=\pi
R$ brane are given by 
\begin{equation}
D_AJ^{a,A}(y)=Q^a(y)\delta(y)~,~~~~~
D_AJ^{a,A}(y)=Q^a(y)\delta(y-\pi R)
\end{equation}
respectively. 
Note that the contributions to the anomaly from the
bulk and brane fermion fields differ by a factor of 2.
So, unless the (rational) ${\cal U}(1)$ charges of all the fermion fields
satisfy
${\rm Tr}_{brane}[T_a\{T_b,T_c\}]=-
\frac{1}{2}{\rm Tr}_{bulk}[T_a\{T_b,T_c\}]$ on both branes, we cannot
cancel
the anomalies induced by the bulk fermions simply by adding brane
fermions. In general, if we insist on rational ${\cal U}(1)$ charges, such
an assignment will not be possible. However, using 
a combination of additional fermion fields and a CS term in the 
action, we can cancel the local gauge anomalies everywhere.

The (nonsupersymmetric) CS action is given by
\begin{equation}
S_{CS}=\int_M
\chi(y){\rm Tr}\left[\A\F^2-
\frac{1}{4}\A^2\F-\frac{1}{4}\F\A^2
\frac{1}{10}\A^4\right]
\la{5DCS}
\end{equation} 
where $\A=A_{\mu }^aT^adx^{\mu }$,
$\F=\fr{1}{2}F_{\mu \nu }^aT^adx^{\mu }\wedge dx^{\nu }$, and
$M$ is the spacetime manifold.
This is a slightly modified form of the CS action because of the
addition of a neutral field $\chi$, which could either be a dynamical
field whose VEV satisfies eq. 
(\ref{profile}) or a nondynamical function.
Since the Lagrangian must be even, $\chi$ has to
have a negative $Z_2$ parity. So, unless $\chi$ is trivially zero
everywhere, it has to have a y-dependence. 

Under an infinitesimal gauge transformation which transforms the fermion
fields, $\psi$, 
\begin{equation}
\psi\rightarrow\psi+i\om \psi ~, 
\end{equation}
we can show that
\begin{eqnarray}
\delta \A=i\om \A-i\A\om +\frac{1}{g}d\om \\
\delta \F=i\om \F-i\F\om \\
\delta S_{\rm CS}=-\frac{1}{g}\int_M d\chi {\rm Tr}[\om \F^2].
\end{eqnarray}
{}From these equations, we have: 
$$
\delta(S_{\rm CS}+S_{\rm rest})=
{\rm Tr}\l \frac{\delta}{\delta
\A}(S_{\rm CS}+S_{\rm rest})\cdot\delta\A\r +(\delta
\psi\frac{\delta}{\delta
\psi}+\delta\phi\frac{\delta}{\delta
\phi}+\ldots)(S_{\rm CS}+S_{\rm rest}) =
$$
\beq
\int d^4xdy
{\rm Tr}\l J^{\mu }\cdot(i[\om ,A_{\mu }]+\frac{1}{g}\pl_{\mu }\om
)\r 
=\int d^4xdy {\rm Tr}\l i J^{\mu }\cdot[\om ,A_{\mu }]-\frac{1}{g}\om
\nabla_{\mu }\cdot J^{\mu }\r~,  
\la{variation}
\eeq 
where $\phi $ represents the charged scalar fields and $\ldots$ represents
the variation due to all the other charged fields. In the previous 
equation, we have split the 
action into three parts, 
$S=S_{\rm CS}+S_{\rm gaugekinetic}+S_{\rm rest}$ and used the 
definition $\J\equiv\frac{\delta}{\delta 
\A}(S_{\rm CS}+S_{\rm rest})$ for the current. $S_{\rm rest}$ 
includes all the terms of the action except the CS and the gauge kinetic 
term and is a functional of $\A$ because the covariant derivative in used 
in 
the matter part of the action. 
But since 
$S_{\rm rest}$ is gauge invariant by assumption, \begin{equation}
D_A
J^{a,A}=\frac{d\chi}{dy}\epsilon^{\alpha\beta\gamma\delta}
F^b_{\alpha\beta}F^c_{\gamma\delta}{\rm Tr}[T^a\{T^b, T^c\}].
\end{equation}
Since we only want 
anomaly cancellation on the fixed points, $\chi$ ought to have the
following profile, 
\begin{equation} 
\chi =\left\{ \begin{array}{ll} 
~~\chi_0,~~~0<y<\pi R~\\
-\chi_0,~~~\pi R<y<2\pi R
\end{array}
\right.~.
\la{profile} 
\eeq
With this form for $\chi $, the 4D anomalous terms induced from variation
of the CS action has
opposite signs on both branes. Now, with the addition of brane fermions
on the $y=0$ brane with the
appropriate quantum numbers to contribute an anomaly of $-Q$, the anomaly
on the $y=0$ brane is $-Q/2$ and the anomaly on the $y=\pi R$ brane is
$Q/2$. But since the anomalies are now of opposite signs, they can be
cancelled by the CS action with an appropriate value for $\chi_0$. Another
possibility is to add the brane fermions to the $y=\pi R$ brane instead.
Now, the anomalies would be $Q/2$ and $-Q/2$ on the $y=0$ and the $y=\pi
R$ branes respectively. This can also be cancelled by the CS action. A
third possibility, of course, is to have the additional fermions in the
bulk, obeying the same $Z_2$ projection as the other fields. The
additional fermion fields would then have chiral zero modes and
massive vector Kaluza-Klein modes from a 4D point of view. In this case,
the anomalies cancel locally and no CS counterterm is needed. But in fact,
however, it can be shown that in the limit as the absolute value of the 5D
mass, $|M|$, of the
additional fermion fields goes to 
infinity\footnote{Under orbifold $Z_2$ parity $M\to -M$, so between an
additional chiral state $\Psi $ and its mirror $\ov{\Psi }$ the coupling
$M\Psi \ov{\Psi }$ is allowed.},
the low energy effective theory
would be that of a chiral brane field plus an effective CS
action \cite{rat} with
the appropriate value for $\chi$ to cancel the anomalies, reducing to the
other two possibilities mentioned earlier.

As far as the mixed anomalies are concerned, for their cancellation we
introduce some
additional superfields. Namely, an $SU(3)$
triplet, $F_1$,  and an $SU(3)$ antitriplet, $F_2$, which are neutral
under
$SU(2)$ and
$U(1)_Y$ (other possibilities seem to give rise to ${\cal U}(1)$ charge
assignments in such a way that either  
suppression of proton decay does not hold, or the additional states obtain
masses of order the electroweak scale). 
These additional fields couple to each other on the brane through the
interaction
term $X^kF_1F_2$, where $-k$ is the sum of the ${\cal U}(1)$ charges of
both fields. 
Referring to Table \ref{t:scalar}, we can see that the mixed  
$SU(3)^2-{\cal U}(1)$, $SU(2)^2-{\cal U}(1)$, $U(1)_Y^2-{\cal U}(1)$ and
$U(1)_Y-{\cal U}(1)^2$ anomalies vanish if the following relations
hold:
$$
\gamma=n+k-\al~,~~~~
b=\alpha-3a-\frac{2}{3}n-\frac{2}{3}k~,
$$
\beq
n=3k~,~~~~\al =\fr{1}{16}(60a+39k)~.
\la{rels}
\eeq
This leaves us with the ${\cal U}(1)^3$ and the 
${\cal U}(1)-{\rm grav}^2$
anomalies. For cancellation of ${\cal U}(1)^3$
anomaly we invoke the bulk CS term. 
For ${\cal U}(1)-{\rm grav}^2$ anomaly
cancellation we add
additional $SU(3)\times SU(2)\times U(1)_Y$ singlet fields which are
charged under ${\cal U}(1)$, such that ${\rm Tr}Q_{{\cal U}(1)}=0$.
The latter condition also avoids  divergences in the renormalization of
the Fayet-Iliopoulos term \cite{nilles}.

\section{Neutrino Masses}

The 4D superpotential couplings which generate the charged fermion masses are given by
\beq
W^{(4)}_Y=qu^ch_u+\l \fr{X}{M_f} \r^nqd^ch_d+
\l \fr{X}{M_f}\r^nle^ch_d~,
\la{4Dyuk}
\eeq
where $M_f$ denotes some fundamental mass scale.
A nonzero VEV for the scalar component of $X$ is guaranteed by a brane
Fayet-Iliopoulos (FI) term for $V_{{\cal U}(1)}$, which is permitted by
all 4D
symmetries. One can also show that within the 5D orbifold framework, the
brane FI term does not induce SUSY
breaking. In appendix A, we present a detailed analysis of these
issues.
We assume that $\lan X\ran $ (${\cal U}(1)$ breaking scale) 
is not
too far
below $M_f$, i.e.
\beq
\fr{\lan X\ran}{M_f}\equiv \ep \simeq 0.2~.
\la{brscale}
\eeq
This value of $\ep $ is an important expansion parameter for 
understanding the charged fermion mass 
hierarchies and 
mixings \cite{anu1}.
Since $\tan \bt \simeq \fr{m_t}{m_b}\ep^n$, $n$
has to take values between $0$ and $3$ to reproduce the observed masses.
Here we consider two scenarios: {\bf (I)} 
$M_f\simeq M_{\rm Pl}=2.4\cdot 10^{18}$~GeV and
{\bf (II)} $M_f\sim 100$~TeV.

For case {\bf (I)}, the Planck scale $d=5$ operators 
$(lh_u)^2/M_{\rm Pl}$ (if permitted) induce neutrino masses
that are much too low to explain the atmospheric
neutrino anomaly via oscillations. To generate neutrino mass 
$\sim 3\cdot 10^{-2}$~eV,
we have to introduce a right handed neutrino state. 
Introduce an MSSM singlet
$N=2$ supermultiplet 
${\bf {\cal N}}_{N=2}=({\cal N},~\ov{\cal N} )$ with ${\cal U}(1)$ charge
$(Q_{\cal N}, -Q_{\cal N})$ and $Z_2$ parity $(+,~-)$.
Then, only ${\cal N}$ will have a zero mode. 
The relevant 4D superpotential
couplings responsible for neutrino masses are
\beq
W^{(4)}_{\nu }=\l \fr{X}{M_{\rm Pl}}\r^ml{\cal N}h_u+
M_{\rm Pl}\l \fr{X}{M_{\rm Pl}}\r^p{\cal N}^{\hs{0.5mm}2}~,
\la{seesaw}
\eeq
where $m$ and $p$ are non-negative integers.
The light neutrino acquires mass of  order of  
$h_u^2/(M_{\rm Pl}\ep^{p-2m})= (10^{-2}-1)$~eV for
$\ep \simeq 0.2$ and $p-2m=5-8$.
This mass scale for the third generation neutrino 
suggest either hierarchical \cite{anu1}, \cite{nuhier} or degenerate
\cite{nudeg}
masses for the neutrinos, if one wants to account for both the
atmospheric and solar neutrino anomalies (see \cite{atm} and
\cite{sol} respectively).

The couplings in eqs. (\ref{seesaw}) and (\ref{rels})
and the prescriptions of Table \ref{t:scalar} give
\beq
a=\fr{203}{36}k-\fr{2}{3}(p-2m)~~~~{\rm and}~~~~
\al =\fr{283}{12}k-\fr{5}{2}(p-2m)~.
\la{aal}
\eeq


{\bf (II)} For a fundamental scale of $M_f\simeq 100$~TeV, 
the situation is quite
different in the neutrino sector. Here, we do not need to introduce right
handed states. The suppression of (Majorana) neutrino masses can be guaranteed
by
${\cal U}(1)$ symmetry. The relevant
4D coupling is
\beq 
W_{\nu}^{(4)}=\l \fr{X}{M_f}\r^{r}\fr{(lh_u)^2}{M_f}~,
\la{supnulow}
\eeq
(where $r$ is a positive integer),
which gives 
$m_{\nu }\simeq h_u^2\ep^{r}/M_f\simeq (0.1-1)$~eV for 
$\ep \simeq 0.2$ and $r=11-13$.
The couplings in eqs. (\ref{supnulow}) and (\ref{rels}) together with the prescriptions
of Table \ref{t:scalar} give
\beq
a=\fr{1}{252}(24r-53k)~~~~{\rm and}~~~~
\al =\fr{1}{168}(60r+277k)~.
\la{aalII}
\eeq 


The couplings in eqs. (\ref{seesaw}) and (\ref{supnulow}) generate neutrino masses consistent
with current atmospheric neutrinos data [$m_{\nu }\sim (0.1-1)$~eV].
An appropriate scale for solar neutrinos can be obtained either
by introducing heavy right handed neutrino states or using 
specific neutrino mass matrices. The latter can be generated if
${\cal U}(1)$ is
applied as a flavor symmetry \cite{anu1}. Indeed, this  can ensure large,
even
maximal mixings between neutrinos \cite{anu1}, explaining both the solar
and
atmospheric neutrino data.

\section{Baryon Number Conservation \\
and Automatic Matter Parity}

It turns out that with suitable ${\cal U}(1)$ charge assignments, it is
very easy
to forbid all dangerous baryon number violating operators and
obtain automatic matter parity. Table \ref{t:ops} list some
matter parity and baryon number violating operators and their 
${\cal U}(1)$ charges for scenarios {\bf (I)} and {\bf (II)}. To
compute the ${\cal U}(1)$ charges of the couplings in the context of
scenario {\bf (I)}, we use relations (\ref{rels}) and (\ref{aal}) and
the prescriptions of Table \ref{t:scalar}, while in the context of scenario
{\bf (II)}, we use
relations (\ref{rels}) and (\ref{aalII}). In scenario {\bf (I)}, 
as can be seen from Table
\ref{t:ops}, the matter parity violating couplings {\bf (i)}-{\bf (iii)}
 are forbidden for $k=1$ (which
gives $\tan \bt \sim $ unity)  
and $p-2m=5-8$ (to get the correct magnitude for the neutrino masses for 
$\ep \simeq 0.2$) because their effective  ${\cal U}(1)$
charges are fractional.  {}For $p-2m=5, 7$, operator {\bf (iv)} 
is allowed with suppressions
$\ep^{16}$ and $\ep^{7}$ respectively which is not relevant
phenomenologically. Baryon number violating $d=5$ operators 
{\bf (v)} and {\bf (vi)} have positive 
${\cal U}(1)$ charges for any positive integer
$k$ and are therefore forbidden. The same applies to the $d=5$ operator 
{\bf (vii)} which violates baryon number.

\begin{table} \caption{${\cal U}(1)$ charges of a few matter parity and
baryon number violating operators for scenarios {\bf (I)} and {\bf (II)}.}

\label{t:ops} 

$$\begin{array}{|c|c|}
\hline
\vspace{-0.4cm}
&\\
\hspace{0.65cm}{\rm Operator}\hspace{0.38cm}&
\hspace{0.5cm}{\rm Corresponding}~{\cal U}(1)~{\rm charge}
~~~\hspace{0.3cm} \\
\vspace{-0.4cm}
&\\
\hline
\end{array}$$   

\vspace{-0.5cm}

 $$\begin{array}{|c|c|c|c|}

\hline
&&{\rm Scenario} ~{\bf (I)}&
{\rm Scenario} ~{\bf (II)}\\
 
\hline 
\vspace{-0.4cm} 
&&& \\
{\bf (i)} & h_ul & -\fr{235}{6}k+\fr{9}{2}(p-2m) &
\fr{53}{168}k-\fr{9}{14}r   \\
\vspace{-0.4cm} 
&&&\\ 
\hline 
\vspace{-0.4cm} 
&&& \\
{\bf (ii)} &qd^cl & -\fr{175}{12}k+2(p-2m) &
\fr{83}{28}k-\fr{2}{7}r     \\
\vspace{-0.4cm} 
&&&\\ 
\hline 
\vspace{-0.4cm} 
&&& \\
{\bf (iii)} &e^cll &-\fr{229}{6}k+\fr{9}{2}(p-2m) &
\fr{221}{168}k-\fr{9}{14}r    \\
\vspace{-0.4cm} 
&&&\\ 

\hline 
\vspace{-0.4cm} 
&&& \\
{\bf (iv)} &u^cd^cd^c &-\fr{77}{2}k+\fr{9}{2}(p-2m) & 
\fr{55}{56}k-\fr{9}{14}r   \\
\vspace{-0.4cm} 
&&&\\ 

\hline 
\vspace{-0.4cm} 
&&& \\
{\bf (v)} &qqql &\fr{4}{3}k & \fr{4}{3}k  \\
\vspace{-0.4cm} 
&&&\\
 
\hline 
\vspace{-0.4cm} 
&&& \\
{\bf (vi)} &u^cu^cd^ce^c &\fr{2}{3}k &  \fr{2}{3}k \\
\vspace{-0.4cm} 
&&&\\

\hline
\vspace{-0.4cm} 
&&& \\ 
{\bf (vii)} &qqqh_d &\fr{73}{2}k-\fr{9}{2}(p-2m) &
-\fr{167}{56}k+\fr{9}{14}r   \\

\hline

\end{array}$$
 
\end{table}

As far as scenario {\bf (II)} is concerned, for $k=1$ and $r=11-13$ 
(which give the correct values for 
the neutrino 
mass (\ref{supnulow}) for $\ep \simeq 0.2$), all {\bf (i)}-{\bf (vii)}
couplings
carry noninteger ${\cal U}(1)$ charges and are therefore
forbidden as a result. Thus, thanks to the ${\cal U}(1)$ 
symmetry, matter parity is present and
baryon number conservation holds, even after taking account of 
dimension five operators.

In scenario {\bf (I)}, higher order baryon and lepton number violating
operators are irrelevant from the phenomenological viewpoint since even
if they are present, they are strongly suppressed by appropriate powers of
$M_{\rm Pl}$.
Therefore, we can conclude that in scenario {\bf (I)}, with the help of
${\cal U}(1)$ symmetry and suitable choices for $a$ and $b$, baryon
number is essentially conserved.

In scenario {\bf (II)}, the situation can be different because of the
low scale of $M_f\simeq 100$~TeV. Operators with $\De B=2$ can induce
observable
processes (such as $n-\ov{n}$ oscillations and  deuteron two body decays
$D\to K^*K$). $\De B=2$ operators of the form
\beq
\fr{1}{M_f^3}u^cd^cd^cu^cd^cd^c~,
\la{B2}
\eeq
have a ${\cal U}(1)$ charge of $-\fr{9}{7}r+\fr{55}{28}k$ (see eqs.
(\ref{rels}) and (\ref{aalII}) and Table \ref{t:scalar}),
which is fractional for $k=1$ and $r=11-13$ and therefore forbidden.
Higher order operators with $\De B \geq 3$ are 
phenomenologically not relevant.

\section{Conclusions}

Throughout our discussion so far, we have assumed flavor independent
${\cal U}(1)$
charges for chiral matter. However, automatic matter parity and baryon
number conservation would hold even if ${\cal U}(1)$ is
regarded
as a flavor symmetry. This provides us with the possibility of explaining the hierarchies between 
the charged fermion masses and the 
CKM matrix elements naturally. Also, one can construct various neutrino oscillation
models in the spirit of ref. \cite{anu1}, accommodating both 
the recent atmospheric and solar neutrino data.

In our considerations the breaking of~ ${\cal U}(1)$ 
symmetry was ensured
by the FI term for $V_{{\cal U}(1)}$ vector superfield. 
An analogous term for 
$V_{U(1)_Y}$ must be avoided in order to avoid 
breaking either SUSY or the SM gauge group in an unacceptable way. 
Note that it will not be induced at the quantum level
because for the MSSM field content we have $Tr[Q_{U(1)_Y}]=0$. Let us 
also note that, since for both scenarios {\bf (I)}, {\bf (II)} the
scale of ${\cal U}(1)$ symmetry breaking lies well above the $Z^0$ boson
mass, the mixed coupling 
$\int d^2\te W_{{\cal U}(1)}W_{U(1)_Y}$ between the field strengths of
${\cal U}(1)$ and $U(1)_Y$ is not dangerous \cite{gaugeB}.

In conclusion, we considered a 5D orbifold construction of $SU(3)_c\tm
SU(2)_L\tm U(1)_Y$
supplemented with an additional ${\cal U}(1)$ gauge factor. This ${\cal U}(1)$ symmetry allows
us to solve various phenomenological puzzles of MSSM, such as baryon number
conservation and the generation of
the desired
neutrino masses for the case where the fundamental scale is
either $M_{Pl}=2.4\cdot 10^{18}$~GeV or
relatively low($\sim 100$~TeV). It turns out that to cancel the mixed and
pure anomalies
arising from the presence of ${\cal U}(1)$, some additional (heavy) states
and 5D Chern-Simons terms must be included. 
The ${\cal U}(1)$ symmetry can also play a role of flavor symmetry for
understanding fermion masses and mixings.

\section*{Appendix A: The Brane FI Term and the Vacuum Structure of the
Fields 
} 
\setcounter{equation}{0}  
\renewcommand{\theequation}{A.\arabic{equation}}

In this appendix we will study the effects of
a brane FI term. The latter gives rise  not only to a non-zero VEV for the
zero mode of
$X$, but also a nonzero VEVs for its KK states and $\Phi^{(k)}$. 
Here, $V$ and $\Phi $ denote the states of  the 5D ${\cal U}(1)$ 
gauge field.
  
The relevant terms for the gauge kinetic type couplings (\ref{lagv}) are 
\beq
{\cal L}_{D}=\fr{1}{g^2}D^2+\fr{1}{g^2}\l -\fr{1}{\sq{2}}\pl_5 D
(\Phi^*+\Phi )+
F^*_{\Phi }F_{\Phi }\r~,
\la{lagD}
\eeq
where in the r.h.s of (\ref{lagD}), the subscript $\Phi$ denotes the component of the 
superfield constructed from $\Phi$ (the same applies for $X$ and
$\ov{X}$).  

The kinetic couplings (\ref{lagf})
for $X$ and $\ov{X}$ (with the ${\cal U}(1)$ charges $Q_X$ and $-Q_X$
respectively), 
\beq
\int d^4\te \l X^{+}e^{Q_XV}X+\ov{X}e^{-Q_XV}\ov{X}^{+}\r +
\int d^2\te \ov{X}\l \pl_5+\fr{Q_X}{\sq{2}}\Phi \r X~,
\la{kinX}
\eeq
are invariant under the gauge transformation
$$
X\to e^{-Q_X\La }X~,~~~~
\ov{X}\to e^{Q_X\La }\ov{X}~,
$$
\beq
V\to V+\La +\La^{+}~,~~~~\Phi \to \Phi +\sq{2}\pl_5 \La ~.
\la{transf}
\eeq
The relevant couplings coming from (\ref{kinX}) are
$$
{\cal L}_{X}=F^*_{X}F_X+F^*_{\ov{X}}F_{\ov{X}}+
F_{\ov{X}}\pl_5X+\ov{X}\pl_5F_X+
$$
\beq
\fr{Q_X}{2}DX^*X-\fr{Q_X}{2}D\ov{X}^*\ov{X}+
\fr{Q_X}{\sq{2}}\l F_{\ov{X}}\Phi X+\ov{X}F_{\Phi }X+
\ov{X}\Phi F_X \r~.
\la{lagX}
\eeq
We also consider a 4D FI-term on a fixed point of the form
\beq
{\cal L}_{\rm FI}^{(4)}=\xi \int d^4\te V=\fr{1}{2}\xi D~,
\la{FI}
\eeq
(${\cal L}_{\rm FI}^{(4)}$ is invariant under the 5D gauge transformation 
$V\to V+\La +\La^{+}$ since $\int d^4\te \La =\int d^4\te \La^{+}=0$).
With the orbifold parities of Table \ref{t:scalar}, we can expand
$X$, $\ov{X}$, $V$ and $\Phi$ as
eq. (\ref{expan1}). Thus,
\beq
V=\sq{2}\sum_{n=0}^{\infty }V^{(n)}\eta^{(n)}\cos \fr{ny}{R}~,~~~
\Phi=\sq{2}\sum_{n=1}^{\infty }\Phi^{(n)}\sin \fr{ny}{R}~.
\la{expans}
\eeq
Substituting in eq. (\ref{lagD}) and integrating
over the fifth coordinate $y$, we obtain
\beq
{\cal L}_{D}=\fr{1}{2g_4^{2}}\sum_{n=0}^{\infty }D^{(n)}D^{(n)}+
\fr{1}{g_4^{2}}\sum_{n=1}^{\infty }\l \fr{n}{R}D^{(n)}\Si^{(n)}+
F^{(n)*}_{\Phi }F^{(n)}_{\Phi }\r~,
\la{kinKK}
\eeq
where
\beq
g_4\equiv \fr{g}{\sq{\pi R}}~.
\la{4Dg}
\eeq
The other terms in eq. (\ref{lagX}) can be expanded to yield
\beq
{\cal L}_X^{(1)-(4)}=\sum_{n=0}^{\infty }F^{(n)*}_{X}F^{(n)}_X+
\sum_{n=1}^{\infty }F^{(n)*}_{\ov{X}}F^{(n)}_{\ov{X}}-
\sum_{n=1}^{\infty }\fr{n}{R}\l F^{(n)}_{\ov{X}}X^{(n)}
+\ov{X}^{(n)}F^{(n)}_X\r~,
\la{lagX14}
\eeq
$$
{\cal L}_X^{(5)}=\fr{Q_X}{2}\left [ 
D^{(0)}\sum_{n=0}^{\infty }X^{(n)*}X^{(n)}+
\fr{1}{\sq{2}}\sum_{n+p\neq 0}D^{(n+p)}X^{(n)*}X^{(p)}
\eta^{(n)}\eta^{(p)}+\right.
$$
\beq
\left.\fr{1}{\sq{2}}\sum_{n\neq p}D^{(|n-p|)}X^{(n)*}X^{(p)}
\eta^{(n)}\eta^{(p)}
 \right ]
\la{lagX5}
\eeq
\beq
{\cal L}_X^{(6)}=\hs{-0.1cm}-\fr{Q_X}{2}\hs{-0.1cm}\left [ 
\hs{-0.1cm}D^{(0)}\sum_{n=1}^{\infty }\ov{X}^{(n)*}\ov{X}^{(n)}
\hs{-0.1cm}+\hs{-0.1cm}
\fr{1}{\sq{2}}\sum_{n\neq p}
\hs{-0.1cm}D^{(|n-p|)}\ov{X}^{(n)*}\ov{X}^{(p)}
\hs{-0.1cm}-\hs{-0.1cm}
\fr{1}{\sq{2}}\sum_{n, p}\hs{-0.1cm}D^{(n+p)}
\ov{X}^{(n)*}\ov{X}^{(p)}
\right ]
\la{lagX6}
\eeq

$$
{\cal L}_X^{(7)}=\fr{Q_X}{\sq{2}}\left [
\sum_{n=1}^{\infty }\l X^{(0)}F^{(n)}_{\ov{X}}\Phi^{(n)}+
X^{(0)}\ov{X}^{(n)}F^{(n)}_{\Phi }+
F^{(0)}_{X}\ov{X}^{(n)}\Phi^{(n)}\r+\right.
$$
$$
\fr{1}{\sq{2}}\sum_{n\neq p}\l
F^{(p)}_{\ov{X}}\Phi^{(n)}X^{(|n-p|)}+
\ov{X}^{(p)}F^{(n)}_{\Phi }X^{(|n-p|)}+
\ov{X}^{(p)}\Phi^{(n)}F^{(|n-p|)}_{X} 
\r-
$$
\beq
\left.
\fr{1}{\sq{2}}\sum_{n+p\neq 0}\l
F^{(p)}_{\ov{X}}\Phi^{(n)}X^{(n+p)}+
\ov{X}^{(p)}F^{(n)}_{\Phi }X^{(n+p)}+
\ov{X}^{(p)}\Phi^{(n)}F^{(n+p)}_{X}
\r
\right ]~.
\la{lagX7}
\eeq

The FI (\ref{FI}) term is allowed  on a brane
\beq
\int dy\de (y){\cal L}_{\rm FI}^{(4)}=\fr{\xi }{2}D^{(0)}+
\fr{\xi }{\sq{2}}\sum_{n=1}^{\infty }D^{(n)}~.
\la{FIKK}
\eeq
{}Using eqs. (\ref{kinKK})-(\ref{FIKK}), the
$D$ and $F$-terms for the zero modes are
\beq
D^{(0)}=-\fr{g_4^2}{2}\l \xi +Q_X\sum_{n=0}^{\infty }X^{(n)*}X^{(n)}-
Q_X\sum_{n=1}^{\infty }\ov{X}^{(n)*}\ov{X}^{(n)}\r~,
\la{D0}
\eeq
\beq
F_X^{(0)*}=-\fr{Q_X}{\sq{2}}\sum_{n=1}^{\infty }\ov{X}^{(n)}\Phi^{(n)}~.
\la{FX0}
\eeq
The $D$ and $F$-terms of the corresponding KK states are
$$
D^{(k)}=-\fr{g_4^2}{2}\left ( \sq{2}\xi+\fr{2k}{g_4^2R}\Si^{(k)}+
\fr{Q_X}{\sq{2}}\sum_{n+p=k}X^{(n)*}X^{(p)}\eta^{(n)}\eta^{(p)}+
\fr{Q_X}{\sq{2}}\sum_{|n-p|=k}X^{(n)*}X^{(p)}\eta^{(n)}\eta^{(p)}
\right.
$$
\beq
\left. -\fr{Q_X}{\sq{2}}\sum_{n+p=k}\ov{X}^{(n)*}\ov{X}^{(p)}-
\fr{Q_X}{\sq{2}}\sum_{|n-p|=k}\ov{X}^{(n)*}\ov{X}^{(p)}
\right ) ~,
\la{Dkk}
\eeq
\beq
F_{X}^{(k)*}=\fr{k}{R}\ov{X}^{(k)}-\fr{Q_X}{2}
\sum_{|n-p|=k}\ov{X}^{(p)}\Phi^{(n)}+
\fr{Q_X}{2}\sum_{n+p=k}\ov{X}^{(p)}\Phi^{(n)}~,
\la{FXkk}
\eeq
\beq
F_{\ov{X}}^{(k)*}=\fr{k}{R}X^{(k)}-\fr{Q_X}{\sq{2}}X^{(0)}\Phi^{(k)}-
\fr{Q_X}{2}\sum_{n\neq k}\Phi^{(n)}X^{(|n-k|)}+
\fr{Q_X}{2}\sum_{n+k\neq 0}\Phi^{(n)}X^{(n+k)}~,
\la{FbarXkk}
\eeq
\beq
F_{\Phi }^{(k)*}=-\fr{g_4^2}{2}Q_X\l \sq{2}X^{(0)}\ov{X}^{(k)}+
\sum_{p\neq k}\ov{X}^{(p)}X^{(|p-k|)}-
\sum_{p+k\neq 0}\ov{X}^{(p)}X^{(p+k)}\r~.
\la{FPhikk}
\eeq

It is easy to see that there is a solution with zero $D$ and $F$
terms and nonzero VEVs for the $X^{(k)}$ and $\Phi^{(k)}$ states.
{}Assuming $\lan \ov{X}^{(k)}\ran =0$ for all $k$, from eqs. (\ref{FX0}), (\ref{FXkk}) and
(\ref{FPhikk}),  we
see that 
\beq
F_X^{(0)}=F_{X}^{(k)}=F_{\Phi }^{(k)}=0~.
\la{vanFs}
\eeq
{}If we require all the other $D$ and $F$-terms to vanish, from eqs. (\ref{D0}),
(\ref{Dkk}) and (\ref{FbarXkk}), we obtain
\beq
\xi +Q_X\sum_{n=0}^{\infty }X^{(n)*}X^{(n)}=0~,
\la{D0van}
\eeq
\beq
\sq{2}\xi+\fr{2k}{g_4^2R}\Si^{(k)}+
\fr{Q_X}{\sq{2}}\sum_{n+p=k}X^{(n)*}X^{(p)}\eta^{(n)}\eta^{(p)}+
\fr{Q_X}{\sq{2}}\sum_{|n-p|=k}X^{(n)*}X^{(p)}\eta^{(n)}\eta^{(p)}=0~,
~~k\neq 0~,
\la{Dkkvan}
\eeq
\beq
\fr{k}{R}X^{(k)}-\fr{Q_X}{\sq{2}}X^{(0)}\Phi^{(k)}-
\fr{Q_X}{2}\sum_{n\neq k}\Phi^{(n)}X^{(|n-k|)}+
\fr{Q_X}{2}\sum_{n+k\neq 0}\Phi^{(n)}X^{(n+k)}=0~,
~~k\neq 0~.
\la{FbarXkkvan}
\eeq
If one assumes that the VEVs of all the $\Phi^{(k)}$ states vanish, then
from eq. (\ref{FbarXkkvan}), we deduce $\lan X^{(k)}\ran =0$ (for $k\neq 0$)
and so, we cannot satisfy eq. (\ref{Dkkvan}). Thus, we can conclude
that, in order to satisfy eqs. (\ref{D0van})-(\ref{FbarXkkvan}) simultaneously,
the states $\Phi^{(k)}$ must have non-zero VEVs.
In order to satisfy eq. (\ref{D0van}), we need opposite signs for 
$\xi $ and $Q_X$. Without any loss of generality, one can assume
$\xi <0$ and $Q_X>0$.
{}If we restrict eqs. (\ref{D0van})-(\ref{FbarXkkvan}) to the first $k$ KK modes of $\Phi$, the 
first $k'$ modes of $X$ and the zero mode
$X^{(0)}$, we are left with $k+k'+1$ nontrivial equations.
Therefore, the number of equations and variables coincides and there will
always
be a solution where all the $D$ and $F$ terms vanish. 
In particular, the $X^{(0)}$ state has a nonzero VEV.

We have shown that within the framework of 5D $S^{(1)}/Z_2$ orbifold
models,
the brane
FI term for the ${\cal U}(1)$ gauge superfield ensures a non-zero VEV for 
the $X$ field and SUSY remains
unbroken.
It turns out that the VEV of the scalar component of $X$ is crucial for
the generation of sufficiently suppressed neutrino masses and to
explain hierarchies between fermion masses and mixings if ${\cal U}(1)$
is applied as a flavor symmetry. 

\vs{0.2cm}

{\bf Acknowledgments}

We would like to thank the Alexander von Humboldt St\"aftung
and NATO Grant PST.CLG.977666 for providing the impetus for this
collaboration. Q.S. also thanks the Theory Group at the University of
Heidelberg, especially Michael Schmidt and Christof Wetterich, for their
hospitality during the final stages of this work. This work is supported
in part by the DOE under contract DE-FG02-91ER 40626.


\bibliographystyle{unsrt}

\end{document}